\documentclass[12pt]{article}
\usepackage{epsfig,graphics,graphicx}
\usepackage{amssymb}
\usepackage{amsmath}
\usepackage{extarrows}
\usepackage[utf8]{inputenc}
\usepackage[english]{babel}
\usepackage{geometry}
\usepackage{amsthm}
\usepackage{scalerel}[2014/03/10]
\usepackage[usestackEOL]{stackengine}
\usepackage{esint}
\usepackage{xcolor}
\usepackage{authblk}
\newtheorem{theo}{Theorem}

\newtheorem{prop}[theo]{Proposition}

\newtheorem{definition}{Definition}
\def\Z{\mathbb Z}

\def\N{\mathbb N}

\def\bw{\mathbf{w}}
\def\bv{\mathbf{v}}
\def\bu{\mathbf{u}}

\def\div{\operatorname{div}}

\title{\sc{Introduction to Number Theoretic Transform}}
\date{}

\title{\sc{Introduction to Number Theoretic Transform}}

\author[1,2]{Banhirup Sengupta}
\author[2]{Peenal Gupta}
\author[3]{Souvik Sengupta}

\affil[1]{Center for Applicable Mathematics, Tata Institute Of Fundamental Research, Bangalore, India}
\affil[2]{Research Group, PinakashieldTech OÜ, Tallinn, Estonia}
\affil[3]{Digital Ecosystems, IONOS SE, Karlsruhe, Germany}

\begin{document}
	
	\maketitle
	
	\abstract{The Number Theoretic Transform (NTT) can be regarded as a variant of the Discrete Fourier Transform. NTT has been quite a powerful mathematical tool in developing Post-Quantum Cryptography and Homomorphic Encryption. The Fourier Transform essentially decomposes a signal into its frequencies. They are traditionally sine or cosine waves. NTT works more over groups or finite fields rather than on a continuous signal and polynomials work as the analog of sine waves in case of NTT. Fast Fourier Trnasform (FFT) style NTT or fast NTT has been proven to be useful in lattice-based cryptography due to its ability to reduce the complexity of polynomial multiplication from quadratic to quasilinear. We have introduced the concepts of cyclic, negacyclic convolutions along with NTT and its inverse and their fast versions.}
	
	\section{Introduction}
	Lattice-based cryptography has emerged as a promising candidate for public-key cryptography which is quantum safe in nature. Many lattice-based schemes are based on operations in the ring of polynomials of the form $\left(\Z/q\Z\right)\left[X\right]/\left(f(X)\right)$, where $f$ is an irreducible polynomial over $\Z$ and $q$ is a prime number. Those schemes, whose security heavily depends on the hardness of the Ring-LWE problem, $f$ is usually chosen to be a exponent of $2$ cyclotomic $X^{n} + 1$, whose roots have order $2n$ and $q$ satisfies the condition $q \equiv 1 \mod{2n}$. The motivation for choosing exponent of $2$ cyclotomic rings and fully splitting primes is that the splitting behavior of such primes in these rings allows for swift multiplication using FFT, which is called NTT when it is performed over the base field $\Z_{q}$. NTT transforms polynomials from the time domain to a frequency (NTT) domain, allowing for coefficient-wise multiplication. Then the inverse NTT is used to transform the output back to the time domain, thereby providing a speed-up for polynomial multiplications. An advantage of NTT-based multiplication over other methods is that NTTs can be saved by directly sampling polynomials in the NTT domain, by storing NTT domain representations of polynomials for later use, and making use of the linearity of NTT when computing sums of products of polynomials as well. This helps in the speed-up as multiplication is one of the most time consuming operation in lattice-based signature schemes such as Dilithium and Falcon. \\ 
	This note is mostly taken from \cite{SML}, \cite{S}, and organized as follows. In Section 2, we give the definitions of cyclic (positive-wrapped) and negacyclic (negative-wrapped) convolutions. Section 3 deals with the positive-wrapped convolution based on NTT, whereas Section 4 is responsible for negative-wrapped convolution. In Section 5, we have discussed about the Cooley-Tukey and Gentleman-Sande algorithms for Fast-NTT and Fast-INTT respectively. Finally, NTT based multiplication has been described in Section 6 along with an example of Dilithium signature scheme.  
	\section{Cyclic and Negacyclic Convolution}
	
	\begin{definition}{\label{Cyclic Convolution}}
		Let $G(x)$ and $H(x)$ be polynomials of degree $n-1$ in the quotient ring $\Z_{q}[x]/(x^{n} - 1)$, where $q\in\Z$. A cyclic convolution or positively wrapped convolution, $PWC(x)$ is defined as:
		$$\aligned
		PWC(x) = \sum_{k=0}^{n-1} c_{k} x^{k}
		\endaligned$$ where $c_{k} = \sum_{i=0}^{k} g_{i} h_{k-i} + \sum_{i=k+1}^{n-1} g_{i} h_{k+n-i} \mod{q}$. If $Y(x)$ is the result of their linear convolution in the ring $\Z_{q}[x]$, it can be defined as well as :
		$$\aligned 
		 PWC(x) = Y(x) \mod{(x^{n} - 1)}.
		 \endaligned$$
	\end{definition}
	\noindent Negacyclic convolution is exactly the same type of convolution, modulo the divisor. The cyclic convolution uses $x^{n} - 1$, while negacyclic convolution uses $x^{n} + 1$. Both these convolution techniques have $O(n^2)$ complexity.
	
	\section{Positive-Wrapped Convolution based on NTT}
	This section describes NTT and its inverse (INTT) based on the $n$-th root of unity, $\omega$.
		\begin{definition}{\label{Primitive nth root of unity}}
		Let $\Z_{q}$ be an integer ring modulo $q$, and $n - 1$ is the polynomial degree of $G(x)$ and $H(x)$. We define $\omega$ as the primitive $n$-th root of unity in $\Z_{q}$ iff :
		$$\omega^{n}\equiv 1 \mod{q}$$ and $$\omega^{k} \not\equiv 1 \mod{q}$$ for $k<n$.
	\end{definition}
	\noindent One must note that the primitive $n$-th root of unity in a ring $\Z_{q}$ might not be unique. For example, in the ring $\Z_{7681}$, $\omega = 3383$ and $\omega = 4298$ are the primitive $4$-th roots of unity.
	\begin{definition}{\label{NTT based on omega}}
		The Number Theoretic Transform (NTT) of a vector with polynomial coefficients $\bv$ is defined as $\hat{\bv} = NTT (\bv)$, where
		$$\aligned
		\hat{\bv}_{j} =\sum_{i=0}^{n-1}\omega^{ij}\bv_{i}\mod{q},
		\endaligned$$ $j=0,1,\dots, n-1$.
	\end{definition}
	\begin{definition}{\label{INTT}}
		The inverse of an NTT vector $\hat{\bv}$ is defined as $\bv = INTT (\hat{\bv})$, where
		$$\aligned
		\bv_{i} = \frac{1}{n}\sum_{j=0}^{n-1}\omega^{-ij}\hat{\bv}_{j}\mod{q},
		\endaligned$$ $j=0,1,\dots, n-1$.
	\end{definition}
	\noindent The only difference between NTT and INTT is $\omega$ replaced by its inverse in $\Z_{q}$ and a weighted product. It should be noted that $\bv = INTT \left(NTT(\bv)\right)$.
	Since NTT is a variant of Discrete Fourier Transform in polynomial ring, one can use the convolution theorem of DFT to calculate positive wrapped convolution, \cite{AB}, \cite{N}.
	\begin{prop}
		Let $\bu$ and $\bv$ be the multiplicands vectors with polynomial coefficients. The positive-wrapped convolution of $\bu$ and $\bv$ is
		\begin{equation}
			\bw = INTT\left(NTT(\bu) \circ NTT(\bv)\right),
		\end{equation} where $\circ$ is an element-wise vector multiplication in $\Z_{q}$.
	\end{prop}
	
	\section{Negative-Wrapped Convolution based on NTT}
	The scope of implementation of positive-wrapped convolution or cyclic convolution is primarily outside the cryptography domain. For example, it is used in Sch\"{o}nhage-Strassen algorithm \cite{SC} for large integer multiplication. However, in the context of Post-Quantum Cryptography and Homomorphic Encryption, the chosen ring is mostly $\Z_{q}[x]/(x^{n} + 1)$ instead of $\Z_{q}[x]/(x^{n} - 1)$. So, we should calculate the polynomial multiplications using negative-wrapped or negacyclic convolution in such rings. \\
	Next, let us define the $2n$-th root of unity which is essential to calculate negacyclic convolution.
	\begin{definition}{\label{2nth root of unity}}
		Let $\Z_{q}$ be an integer ring modulo $q$, $G(x)$ and $H(x)$ be $n-1$ degree polynomials, and $\omega$ is its primitive $n$-th root of unity. We define $\psi$ as the primitive $2n$-th root of unity iff :
		$$\psi^{2} \equiv \omega\mod{q}$$ and $$\psi^{n} \equiv -1\mod{q}.$$
	\end{definition} 
	\noindent For example, in a ring $\Z_{7681}$ and $n=4$, when $\omega=3383$,  value of $\psi$ can be either $1925$ or $5756$.
		\begin{definition}{\label{Negacyclic NTT}}
		The Negative-Wrapped Number Theoretic Transform of a vector with polynomial coefficients $\bv$ is defined as $\hat{\bv} = NTT^{\psi} (\bv)$, where
		$$\aligned
		\hat{\bv}_{j} =\sum_{i=0}^{n-1}\psi^{i}\omega^{ij}\bv_{i}\mod{q},
		\endaligned$$ $j=0,1,\dots, n-1$. Since $\psi^{2} \equiv \omega\mod{q}$, one can substitute $\omega=\psi^{2}$ above :
		\begin{equation}\label{NWNTT}
				\hat{\bv}_{j} =\sum_{i=0}^{n-1}\psi^{2ij + i}\bv_{i}\mod{q}.
		\end{equation}
	\end{definition}
	\begin{definition}{\label{INTTPSI}}
		Negative-Wrapped inverse of an NTT vector $\hat{\bv}$ is defined as $\bv = INTT^{\psi^{-1}} (\hat{\bv})$ :
		$$\aligned
		\bv_{i} = \frac{1}{n}\sum_{j=0}^{n-1}\psi^{-j}\omega^{-ij}\hat{\bv}_{j}\mod{q},
		\endaligned$$ $j=0,1,\dots, n-1$. Substituting $\omega = \psi^{2}$, we get
		\begin{equation}\label{NWINTT}
			\bv_{i} = \frac{1}{n}\sum_{j=0}^{n-1}\psi^{-(2ij + j)}\hat{\bv}_{j}\mod{q}.
		\end{equation}
	\end{definition}
		\begin{prop}
		Let $\bu$ and $\bv$ be the multiplicands vectors with polynomial coefficients. The negative-wrapped convolution of $\bu$ and $\bv$ is
		\begin{equation}
			\bw = INTT^{\psi^{-1}}\left(NTT^{\psi}(\bu) \circ NTT^{\psi}(\bv)\right),
		\end{equation} where $\circ$ is an element-wise vector multiplication in $\Z_{q}$.
	\end{prop}
	\noindent The modulus $q$ needs to satisfy the following to make NTT transformation possible :
	\begin{itemize}
		\item Then $n$-th root of unity $\omega$ exists in the ring $\Z_{q}$, so that one can perform positive-wrapped convolutions.
		\item The $2n$-th root of unity $\psi$ exists in the ring $\Z_{q}$ to make negative-wrapped convolutions work. The modulus $q$ has to satisfy the following theorems to make sure that $\omega$ and $\psi$ exist respectively.
	\end{itemize}
	The modulus $q$ has to satisfy the following theorem to ensure the existence of $\omega$ \cite{AB}, \cite{P}, \cite{DCD} : 
		\begin{theo}
			If $q$ is prime, then $n$ must $q-1$. If $q$ is composite such that :
			$$\aligned
			q=q_{1}^{m_1}\cdot q_{2}^{m_2}\cdots q_{k}^{m_k}
			\endaligned$$ then $n$ must divide the GCD of $(q_{1}-1, q_{2}-1,\cdots, q_{k}-1)$.
		\end{theo}
		\noindent However, the preceding theorem does not guarantee the existence of $\psi$ in $\Z_{q}$. The next theorem will ensure that $\psi$ exists in $\Z_{q}$. 
		\begin{theo}
			If $q$ is prime, then $2n$ must $q-1$. If $q$ is composite such that :
			$$\aligned
			q=q_{1}^{m_1}\cdot q_{2}^{m_2}\cdots q_{k}^{m_k}
			\endaligned$$ then $2n$ must divide the GCD of $(q_{1}-1, q_{2}-1,\cdots, q_{k}-1)$.
		\end{theo}
		\begin{definition}
			A PWC-NTT friendly modulus $q$ is defined iff an $n$-th root of unity, $\omega$ exists in $\Z_{q}$.
		\end{definition}
				\begin{definition}
			A NWC-NTT friendly modulus $q$ is defined iff an $n$-th root of unity, $\omega$ and $2n$-th root of unity $\psi$ both exist in $\Z_{q}$.
		\end{definition}

    \section{FFT-style NTT}
	Usually NTT has $O(n^2)$ complexity, thereby making no difference from that of negacyclic convolution. However, NTT is Discrete Fourier transform in ring of polynomials. So, DFT optimization techniques can be applied to NTT as well. The popular technique of DFT optimization is called Fast Fourier transform. It was proposed independently by Cooley-Tukey \cite{CT} and Gentleman-Sande \cite{GS}. Both of these methods use similar butterflies divide-and-conquer technique to achieve the quasilinear complexity $O(n\log n)$. One can use divide-and-conquer techniques to fasten the process of matrix multiplication needed for NTT by utilizing the periodicity and symmetry property of $\psi$ : 
	$$\aligned
	\psi^{k+2n} = \psi^{k}
	\endaligned$$ and
	$$\aligned
	\psi^{k+n} = -\psi^{k}.
	\endaligned$$
	\subsection{Cooley-Tukey (CT) Algorithm for Fast-NTT :} 
	The summation in equation \eqref{NWNTT} can be separated into two parts :
	$$\aligned
			\hat{\bv}_{j} &=\sum_{i=0}^{n-1}\psi^{2ij + i}\bv_{i}\mod{q} \\
		&=\sum_{i=0}^{n/2-1}\psi^{4ij + 2i}\bv_{2i} + \sum_{i=0}^{n/2-1}\psi^{4ij + 2j + 2i + 1}\bv_{2i+1}\mod{q} \\
		&=\sum_{i=0}^{n/2-1}\psi^{4ij + 2i}\bv_{2i} + \psi^{2ij+1}\sum_{i=0}^{n/2-1}\psi^{4ij + 2i}\bv_{2i+1}\mod{q}
	\endaligned$$
	 Using $\psi$'s symmetry properties :
	 $$\aligned
	 \hat{\bv}_{j+n/2} = \sum_{i=0}^{n/2-1}\psi^{4ij + 2i}\bv_{2i} - \psi^{2ij+1}\sum_{i=0}^{n/2-1}\psi^{4ij + 2i}\bv_{2i+1}\mod{q}
	 \endaligned$$
	 Let $A_{j} = \sum_{i=0}^{\frac{n}{2}-1}\psi^{4ij + 2i}\bv_{2i}$ and $B_{j} = \sum_{i=0}^{\frac{n}{2}-1}\psi^{4ij + 2i}\bv_{2i + 1}$. Then the above equations become
	 $$\aligned
	 \hat{\bv}_{j} = A_{j} + \psi^{2j + 1} B_{j}\mod{q},
	 \endaligned$$ and 
	 $$\aligned
	  \hat{\bv}_{j + \frac{n}{2}} = A_{j} - \psi^{2j + 1} B_{j}\mod{q}.
	  \endaligned$$
	  Here, $A_{j}$ and $B_{j}$ can be obtained as $\frac{n}{2}$ points NTT. It should be noted that the process can be repeated for all the coefficients if $n$ is of exponent-of-two. The above two equar=tions are together called the \textit{CT butterfly}. 
	  \subsection{Gentleman-Sande (GS) Algorithm for Fast-INTT :}
	  Neglecting the weight $\frac{1}{n}$ in the equation \eqref{NWINTT},
	  $$\aligned
	  			\bv_{i} &= \sum_{j=0}^{n-1}\psi^{-(2i + 1)j}\hat{\bv}_{j}\mod{q} \\
	  			        &= \left[\sum_{j=0}^{\frac{n}{2}-1}\psi^{-(2i + 1)j}\hat{\bv}_{j} + \sum_{j=\frac{n}{2}}^{n-1}\psi^{-(2i + 1)(j + \frac{n}{2})}\hat{\bv}_{j + \frac{n}{2}}\right]\mod{q} \\
	  			        &= \psi^{-i}\left[\sum_{j=0}^{\frac{n}{2}-1}\psi^{-2ij}\hat{\bv}_{j} + \sum_{j=\frac{n}{2}}^{n-1}\psi^{-2i(j + \frac{n}{2})}\hat{\bv}_{j + \frac{n}{2}}\right]\mod{q}.
	  \endaligned$$			        
	  Next, using the symmetry and periodicity of $\psi^{-1}$, we get for the even term :
	  $$\aligned
	  \bv_{2i} &= \psi^{-2i}\left[\sum_{j=0}^{\frac{n}{2}-1}\psi^{-4ij}\hat{\bv}_{j} + \sum_{j=\frac{n}{2}}^{n-1}\psi^{-4i(j + \frac{n}{2})}\hat{\bv}_{j + \frac{n}{2}}\right]\mod{q}\\
	           &= \psi^{-2i}\sum_{j=0}^{\frac{n}{2}-1}\left[\hat{\bv}_{j} + \hat{\bv}_{j + \frac{n}{2}}\right]\psi^{-4ij}\mod{q}.
	  \endaligned$$
	  It is easy to check that using the same derivation for the odd term gives :
	  $$\aligned
	  \bv_{2i+1} &= \psi^{-2i}\sum_{j=0}^{\frac{n}{2}-1}\left[\hat{\bv}_{j} - \hat{\bv}_{j + \frac{n}{2}}\right]\psi^{-4ij}\mod{q}.
	  \endaligned$$
	  Let $A_{i} = \sum_{j=0}^{\frac{n}{2}-1}\hat{\bv}_{j}\psi^{-4ij}$ and $B_{i} = \sum_{j=0}^{\frac{n}{2}-1}\hat{\bv}_{j + \frac{n}{2}}\psi^{-4ij}$. Then we have,
	  $$\aligned
	  \bv_{2i} = \left(A_{i} + B_{i}\right)\psi^{-2i}\mod{q}
	  \endaligned$$ and 
	  $$\aligned
	  \bv_{2i+1} = \left(A_{i} - B_{i}\right)\psi^{-2i}\mod{q}.
	  \endaligned$$
	  Above, $A_{i}$ and $B_{i}$ can be obtained as $\frac{n}{2}$ points INTT. It is clear that the process can be repeated for all the coefficients if $n$ is exponent-of-two. The above two equations are together known as the \textit{GS butterfly}. \\
	  To carry out polynomial multiplication, one can use CT butterflies to transform both inputs to the NTT domain. Then element-wise multiplication is performed to achieve the outputs. The outcome is inverted back using GS butterflies performing INTT. The butterflies play a major role in reducing the complexity of the polynomial multiplication from quadratic to quasilinear. The larger the degree of the polynomial, the greater the speed and minimal cost \cite{H}. 
	  \subsection{Normal and Bit-Reversed Order}
	  \begin{definition}
	  	Let $n$ be an exponent of 2, and $b$ be a non-negative integer such that $b<n$. The bit-reversal of $b$ is defined as :
	  	$$\aligned
	  	brv_{n}\left(b_{\log{n}-1}2^{\log{n}-1} + \cdots + b_{1}2 + b_{0}\right) = b_{0}2^{\log{n}-1} + \cdots + b_{\log{n}-2}2 + b_{\log{n}-1},
	  	\endaligned$$ where $b_i$ is the $i$-th bit of the binary expansion of $b$ \cite{ZY}.
	  \end{definition}
	  \noindent The input of CT Butterfly is in Normal Order (NO) and the output is in Bit-Reversed Order (BO). To the contrary, the input of GS Butterfly is in BO, and the output in NO. However, one can reconfigure the CT butterfly to have BO-input and NO-output, and GS butterfly to have NO-input and BO-output. Usually normal order as NTT input is called decimation in time, whereas bit-reversed order input is called decimation in frequency \cite{SA}. 
	  \section{NTT based multiplication}
	  We describe NTT-based multiplication with an example explaining the reduction of the computation complexity of the polynomial multiplication. \\
	  Let $q$ be prime such that $q\equiv 1\mod{2n}
	  $. So, $2n$ divides the order $q-1$ of the cyclic group $\Z_{q}^{\times}$ (Unit group of $\Z_{q}$). Thus, $\Z_{q}$ contains $n$ primitive $2n$-th roots of unity $\psi^{i}$, where $i = 1,3,\cdots, 2n-1.$ It follows that $x^{n} + 1$ factors into linear polynomials $x - \psi^{i}$ over $\Z_{q}$. In fact, the Chinese Remainder Theorem states that the natural ring homomorphism 
	  $$\aligned
	  f \mapsto \left(f(\psi), f(\psi^3), \cdots, f(\psi^{2n-1})\right) : \Z_{q}[x]/(x^{n} + 1) \to \prod_{i} \Z_{q}[x]/(x - \psi^{i})
	  \endaligned$$ is an isomorphism. The NTT computes this isomorphism and one can write NTT as a mapping from $R_{q}$ to $\Z_{q}^{n}$. Then the product $fg$ of two polynomials $f,g\in R_{q}$ can be calculated as $$fg = INTT\left(NTT(f)NTT(g)\right).$$ This method involves two NTTs, one INTT and the pointwise multiplication in $\Z_{q}^{n}$. However, one can save NTTs. For example,
	  	\begin{equation}
	  	\sum_{i=1}^{t} f_{i}g_{i} = \sum_{i=1}^{t} INTT\left(NTT(f_{i})NTT(g_{i})\right) = INTT\left(\sum_{i=1}^{t}NTT(f_{i})NTT(g_{i})\right)
	  \end{equation}
	  These sums of products of polynomials need to be computed in the matrix vector multiplication of schemes relying on the Module-LWE problem. As an example, we consider the Dilithium signature scheme with the parameters of the highest level security where the matrix $A$ has dimensions $6\times5$. One can save $30$ NTTs by sampling directly in the NTT representation, as $A$ is sampled uniformly random. Then the vector needs to be transformed only once, saving another $25$ NTTs. Now, we need only $6$ INTTs instead of 30 INTTs, because of the linearity of NTT. So, instead of $90$ NTTs, one needs only 11, 5 NTTs to transform the vector and $6$ INTTs.
	  
	  \section{Conclusion}
	  Achieving the quasilinear computational complexity from quadratic in the case of multiplication of polynomials with very high degrees is quite a big challenge. This has been achieved by NTTs. $O(n^2$) time complexity is usually achieved when an alogrithm involves nested loops, where the inner loop iterates through all or a major portion of the input for each iteration of outer loop. The growth of $n^2$ is very rapid for large $n$. This makes such algorithms inefficient for large datasets. Traditional cyclic and negacyclic convolutions have $O(n^2)$ complexity. However, NTTs achieve $O(n\log n)$ time complexity due to its divide and conquer technique. In this type of algorithm, a problem is repeatedly divided into smaller subproblems, all of which are solved first and then combined, thereby bypassing the creation of nested loops, which would make the process relatively slower attaining quadratic complexity. Modern day cryptography, more precisely, lattice-based cryptography is heavily dependent on NTT for efficient polynomial multiplication, which is a core operation for these algorithms. NTTs are the sole reason to get it done in $O(n\log n)$ time instead of $O(n^2)$. NTTs are widely used in PQC algorithms such as Kyber and Falcon among others.

	 \end{document}